\documentclass[twocolumn,aps,showpacs]{revtex4}
\usepackage{epsfig}
\usepackage{graphicx}
\usepackage{dcolumn}

\begin{document}

\title{Hindrance of Heavy-ion Fusion at Extreme Sub-Barrier Energies in 
Open-shell Colliding Systems}

\author{C.~L. Jiang}
\author{K.~E. Rehm}
\author{H. Esbensen}
\author{R.~V.~F. Janssens}
\author{B.~B. Back}\affiliation{Physics Division, Argonne National
Laboratory, Argonne, IL 60439}
\author{P. Collon}\affiliation{University of Notre Dame, Notre Dame, IN 46556}
\author{C.~N. Davids}
\author{J.~P. Greene}
\author{D.~J. Henderson}
\author{C.~J. Lister}\affiliation{Physics Division, Argonne National
Laboratory, Argonne, IL 60439}
\author{S. Kurtz}\affiliation{University of Notre Dame, Notre Dame, IN 46556}
\author{R.~C. Pardo}
\author{T.~Pennington}  \altaffiliation{Deceased} \affiliation{Physics 
  Division, Argonne National Laboratory, Argonne, IL 60439}
\author{M. Paul}\affiliation{Hebrew University, Jerusalem 91904, Israel}
\author{D. Peterson}
\author{D. Seweryniak}
\author{B. Shumard}
\author{S. Sinha}
\author{X.~D. Tang}
\author{I. Tanihata}
\author{S. Zhu}\affiliation{Physics Division, Argonne National Laboratory,
Argonne, IL 60439}

\date{\today}

\begin{abstract}
The excitation function for the fusion-evaporation reaction $^{64}$Ni + $^{100}
$Mo has been measured down to a cross-section of $\sim$ 5 nb. Extensive 
coupled-channels calculations have been performed, which cannot reproduce the steep fall-off of the excitation function at extreme sub-barrier energies. Thus, this system exhibits a hindrance for fusion, a phenomenon that has been discovered only recently. In the $S$-factor representation introduced to quantify the hindrance, a maximum is observed at $E_s$=120.6 MeV, which corresponds to 90\% of the reference energy $E_s^{ref}$, a value 
expected from systematics of closed-shell systems. A systematic analysis of Ni-induced fusion reactions leading to compound nuclei with mass A=100-200 is presented in order to explore a possible dependence of the fusion hindrance on nuclear structure.

\end{abstract}
\pacs{25.70.Jj, 24.10.Eq}
\maketitle

\section{Introduction} 
Heavy-ion induced fusion reactions have been  studied extensively for more than 
forty years, especially since the discovery of the sub-barrier enhancement 
phenomenon \cite {sum1,sum2, sum3, sum4, sum5}. Coupled-channels descriptions 
were shown to explain the phenomenon successfully \cite {hagi0, rowl}. 
Recently, evidence was found for a strong hindrance of the fusion process at 
extreme sub-barrier energies, an effect for which there is no satisfactory 
explanation in present model calculations \cite {jiang0}. A systematic 
survey of existing data from the literature shows that there is a regular 
pattern to the energy dependence for the appearance of fusion hindrance \cite{jiang1}. For stiff, closed-shell 
colliding systems, a significant maximum in the $S$-factor is present as a function of the beam energy. This maximum signals the onset of sub-barrier hindrance, and the energy, $E_s$, at which it is located can be described well by an empirical formula \cite{jiang1}:
\begin{equation}
\label{eq1}
E_s^{\it ref} = 0.356\left(Z_1Z_2\sqrt{\mu}\right)^{\frac{2}{3}} \rm{(MeV)},
\end{equation}
where $\mu=A_1A_2/(A_1+A_2)$. For softer systems, this formula provides an 
upper limit for the energy 
at which the $S$-factor has its maximum.

Fusion hindrance at extreme sub-barrier energies is relevant, not only for 
understanding the dynamics of reactions between complex systems, but also 
for astrophysics and for the synthesis of superheavy elements.

The influence of nuclear structure on this hindrance behavior was first studied
in a detailed comparison \cite {jiang2} for the colliding systems:  
$^{58}$Ni + $^{58}$Ni \cite{beck1},
$^{58}$Ni + $^{60}$Ni \cite{stef3},
$^{58}$Ni + $^{64}$Ni \cite{beck2}, and
$^{64}$Ni + $^{64}$Ni \cite{jiang2}, where the systems are arranged in order of decreasing stiffness. For $^{64}$Ni + $^{64}$Ni, an open-shell colliding system, the measured  location in energy of the $S$-factor maximum is about $9\%$ lower than the 
value expected from Eq.~\ref{eq1} \cite{jiang2}, while the other systems are more 
in line with the systematics.



The aim of the present paper is to further investigate the hindrance 
phenomenon by measuring fusion evaporation for the open-shell system 
$^{64}$Ni + $^{100}$Mo. Compared to $^{64}$Ni, $^{100}$Mo is a transitional 
nucleus with
two protons outside the closed proton shell. Two earlier measurements of 
fusion excitation functions for $^{64}$Ni + $^{100}$Mo can be found in the 
literature \cite{rehm, halb}. The minimum cross section measured in these two 
experiments is about 0.4 mb. Since the hindrance behavior is expected 
(from Eq.~\ref{eq1}) to occur
at much lower energies, the main aim of the present measurement was an
extension of the excitation function into the nb region in order to localize and quantify
this hindrance.

\section{Experimental Procedure and Results} 

The experiment was performed with $^{64}$Ni beams in the energy range of
196-262 MeV from the superconducting linear accelerator ATLAS at Argonne National Laboratory. The maximum beam current used was $\sim$ 
60 pnA. The high melting point of the target material, metallic molybdenum 
evaporated on a 40$\mu$g/cm$^2$ carbon foil, prevented damage 
to the target by the relatively high beam current. The target thickness was 
constant during 
the experiment, as monitored with Si detectors. Thin targets with 
thicknesses of 8 or 18$\mu$g/cm$^2$ were used in order to 
reduce the correction for target thickness in the energy regime of steep 
fall-off of the excitation function. The isotopic abundance of $^{100}$Mo was 
97.42$\%$, with the remainder coming from $^{98}$Mo (0.96\%), $^{97}$Mo 
(0.28\%), $^{96}$Mo (0.34\%), $^{95}$Mo (0.29\%), $^{94}$Mo (0.18\%), and 
$^{92}$Mo (0.53\%). The selection of the beam and target combination excluded 
the possibility of background from fusion reactions coming from beam or target
contaminants~\cite{jiang0,jiang1}. Two surface-barrier Si detectors, located 
at $\pm$ 43$^{\circ}$ with respect to the beam direction, served as monitors. 
The absolute cross sections for fusion-evaporation were determined by using 
elastic scattering measured with the monitors. 

The evaporation residues were identified and measured with the Fragment 
Mass Analyzer (FMA) \cite{dav1}, which has been upgraded with the installation 
of a split-anode in the first electric-dipole \cite{dav3}. The background, 
originating mostly from scattered beam, was greatly suppressed after this 
upgrade. In order to push the cross section measurements to the lowest level, 
a new focal-plane detector with the configuration PGAC-TIC-PGAC-TIC-PGAC-IC,
was used in the experiment. Here, the symbols PGAC stand for $x$-$y$ position 
sensitive parallel grid avalanche counters \cite{hend}, TIC for transmission 
ionization chambers \cite{penn}, and IC for a large volume multi-anode 
ionization chamber. The first PGAC was mounted at the horizontal 
($x$-direction) focal-plane of the FMA.

The $y$-focus condition occurs about 70 cm downstream of the focal-plane, which
is nearly at the middle of the last ionization chamber. The three sets of
position signals $x_1, y_1, x_2, y_2, x_3$ and $y_3$ were measured with the
three PGAC's. The flight-times $t_2$ and $t_3$ obtained from
PGAC$_1$-PGAC$_2$ and PGAC$_1$-PGAC$_3$ 
were also recorded. Seven $\Delta E$ signals were measured with 
the ionization chambers (the first four $\Delta E_1$-$\Delta E_4$ in the two 
transmission ionization chambers and the last three $\Delta E_5$-$\Delta E_7$ 
in the final ionization chamber, IC). Another time-of-flight signal, $t_{rf}$, 
measured the time difference between the rf-system of the accelerator and the 
first PGAC. Three additional $\Delta E$ signals from the PGAC's were also
recorded. The three PGAC's and two TIC's operated  
at a fixed pressure of 3 torr of isobutane. The last ionization chamber had an 
adjustable pressure of 22 - 30 torr of isobutane. There was one 0.13 mg/cm$^2$ 
Mylar pressure foil
in front of the first PGAC, and another one with a thickness of
0.22 mg/cm$^2$ located between the third PGAC and the last ionization chamber. 
This setup allowed for full tracking of each particle detected in the system, 
and provided very good separation between evaporation residues and background 
events. A more detailed description of the new detector system will be 
published elsewhere \cite{jiang3}.  

\begin{figure}[hbt] 
\epsfig{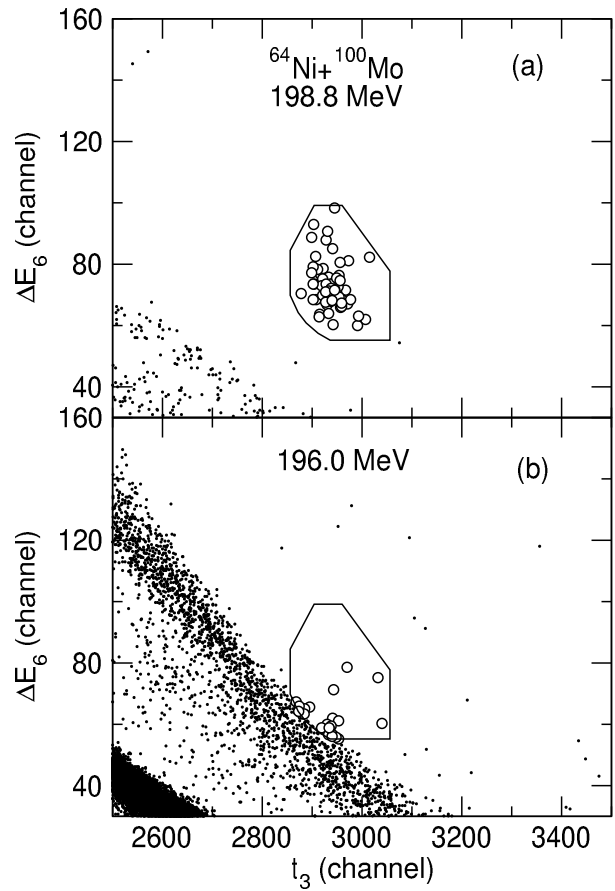}
\caption{
Two-dimensional plots of $\Delta E_6$ vs. $t_3$ for the new focal plane
detector system at the FMA, obtained at sub-barrier incident energies of 198.8 
MeV (a) and 196.0 MeV (b). The isolated group in (a) (open circles)
originates from evaporation residues, whereas the other events are caused by 
background. At E=196.0 MeV (b), 23 events (open circles) fall inside the acceptance window for fusion-evaporation.}
\label{fig2}
\end{figure}

For most settings of the FMA, two charge states of the residues were collected 
simultaneously. For the energies $E_{lab}$ = 260.5, 245.8,
209.1, 207.1 and 202.2 MeV, full charge state distributions were measured,
while for most other energies, two FMA settings, {\it i.e.} four charge 
states were recorded. At the four lowest energies, only two charge states 
were measured. From the full charge state distributions, charge state 
fractions were determined for extrapolation to all other energies. The  
energy distributions and angular distributions of evaporation residues were 
calculated with 
the statistical code PACE \cite{pace}. Total angular distributions for 
fusion-evaporation have been measured in Ref. \cite{rehm} for 
$^{64}$Ni + $^{100}$Mo. In order to check the PACE code, calculations were 
compared with these experimental angular distributions of Ref.~\cite{rehm} and 
good agreement was  found as long as the total calculated angular 
distributions were taken as a weighted sum of the angular distributions of the
different masses from our $m/q$ measurements and folded with multiple scattering. Whereas in Ref. \cite{rehm} rather thick targets were used, the corrections from multiple scattering are small in the present experiment. 
The transport efficiencies of the FMA were calculated with these 
angular distributions together with Monte Carlo simulations, 
using a modified version of the GIOS Code \cite{jian}.
The large momentum acceptance, $\pm$ 10 $\%$,
and the large angular acceptance, $\theta_{lab} <$ 2.3$^\circ$, of the FMA
result in a high detection efficiency for the residues.

\begin{figure}[hbt] 
\epsfig{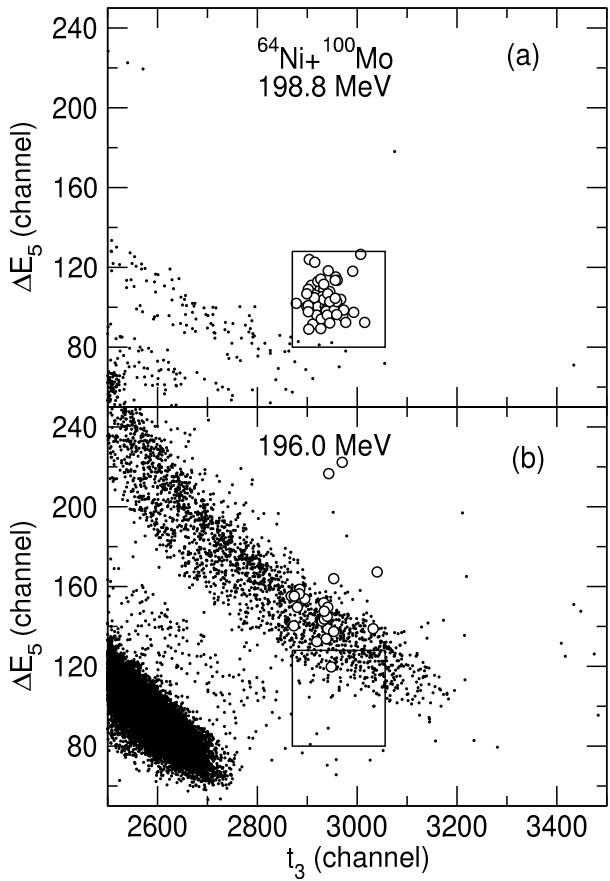}
\caption{
Two-dimensional plots of $\Delta E_5$ vs. $t_3$ for the focal plane
detector system at the FMA, obtained at sub-barrier incident energies of 198.8 
MeV (a) and 196.0 MeV (b). The isolated group in (a) (open circles) 
originates from evaporation residues. The other events are caused by 
background. At 196.0 MeV, only one of the candidate evaporation residue events falls within the $\Delta E_5$ range. See text for details.}
\label{fig4}
\end{figure}

The flight-time $t_3$ between PGAC$_1$ and PGAC$_3$, together with a 
$\Delta E$ signal measured in a counter far behind the focal plane (e.g. $\Delta E_6$), was
found to give the best separation of the residues from background events. 
Two-dimensional plots of $\Delta E_6$ vs. $t_3$ are shown in Fig.~\ref{fig2}, 
indicating the excellent separation achieved with these two quantities. 
Down to a cross section level of $\sim$ 300 nb, the evaporation residues could 
be identified on the basis of these two parameters alone. The events that fall 
within the expected window for evaporation residues are shown as open circles 
in Fig.~\ref{fig2}a. Events shown by small points originate from beam particles, scattered from the beam pipe or from the area of the ``beam stop'' at the first split anode. The events in Fig.~\ref{fig2}a correspond to a cross section of 242 nb obtained in a 12 hours run. 
A two-dimensional plot of $\Delta E_5$ vs. $t_3$, corresponding to the same 
events is shown in Fig.~\ref{fig4}a. At this energy evaporation residues 
(open circles) are again well separated from background events (small points). 

At the lowest energy, $E_{beam}$=196.0 MeV, we find that 23 events fall within the expected evaporation residue window as illustrated in Fig.~\ref{fig2}b. Most of these events are suspiciously close to an intense band of scattered beam particles which borders the fusion evaporation residue window. By examining the $\Delta E_5$ vs. $t_3$ spectrum shown in Fig.~\ref{fig4}b, it is evident that only one of these events falls within the expected range of the $\Delta E_5$ signal. However, this event does not have the correct $m/q$-value as will be demonstrated below.

\begin{figure}[hbt] 
\epsfig{file=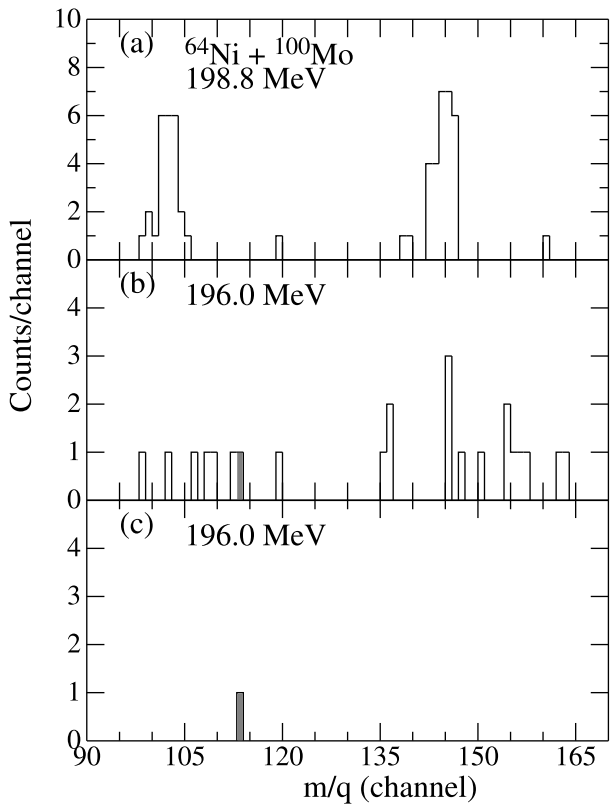,width=8.0cm}
\caption{Position ($m/q$) spectra obtained at energies of 198.8 MeV (a) and 
196.0 MeV (b). These events fall inside the windows in Fig.~\ref{fig2}a or \ref{fig2}b, respectively. The event that also satisfies the $\Delta E_5$ condition is marked solid grey in panels b and c. See text for details.}
\label{fig3}
\end{figure}

An $m/q$ spectrum containing the 23 events that fall inside the $\Delta E_6-t_3$ window is given in Fig.~\ref{fig3}b. This spectrum is spread out over the whole $m/q$ range and does not appear to be associated with fusion evaporation residue events seen at higher energies to be concentrated near channels 103 and 145 (see Fig.~\ref{fig3}a). The only candidate event at $E$=196.0 MeV, which satisfies the conditions for the $\Delta E_5,\Delta E_6$, and $t_3$ signals, appears in channel 114 as seen in Fig.~\ref{fig3}b. By further checking all the detector signals for this event, it was concluded that it is a background event, probably arising from scatterings of a beam particle at several locations through the spectrometer. Hence, only an upper limit for the measured cross section could be determined at this beam energy. These results show that a background suppression factor of about 4$\times$10$^{-17}$ can be achieved with the upgraded FMA and the present detector system.

The uncertainties in the reported evaporation cross sections arise 
mainly from corrections for the charged state distribution, the FMA transport 
efficiency, the detector efficiency, and counting statistics. The total 
uncertainties for the evaporation cross sections are around 10-17$\%$, except 
for the measurements at the two lowest energies where only upper limits can 
be given. The cross sections are listed in Table I.

The fusion-fission cross sections for $^{64}$Ni + $^{100}$Mo have not been
measured previously. They were, however, calculated in Ref. \cite{rehm} with the code CASCADE \cite {puhl}. Similar 
calculations with the same parameters were performed for the present 
experiment. The total fusion cross sections are also listed in Table I, 
together with the fusion-fission cross sections. Rather large uncertainties were
given for the fusion-fission contributions resulting in somewhat larger errors 
for the total cross sections at the highest beam energies.  

\begin{figure}[hbt] 
\epsfig{file=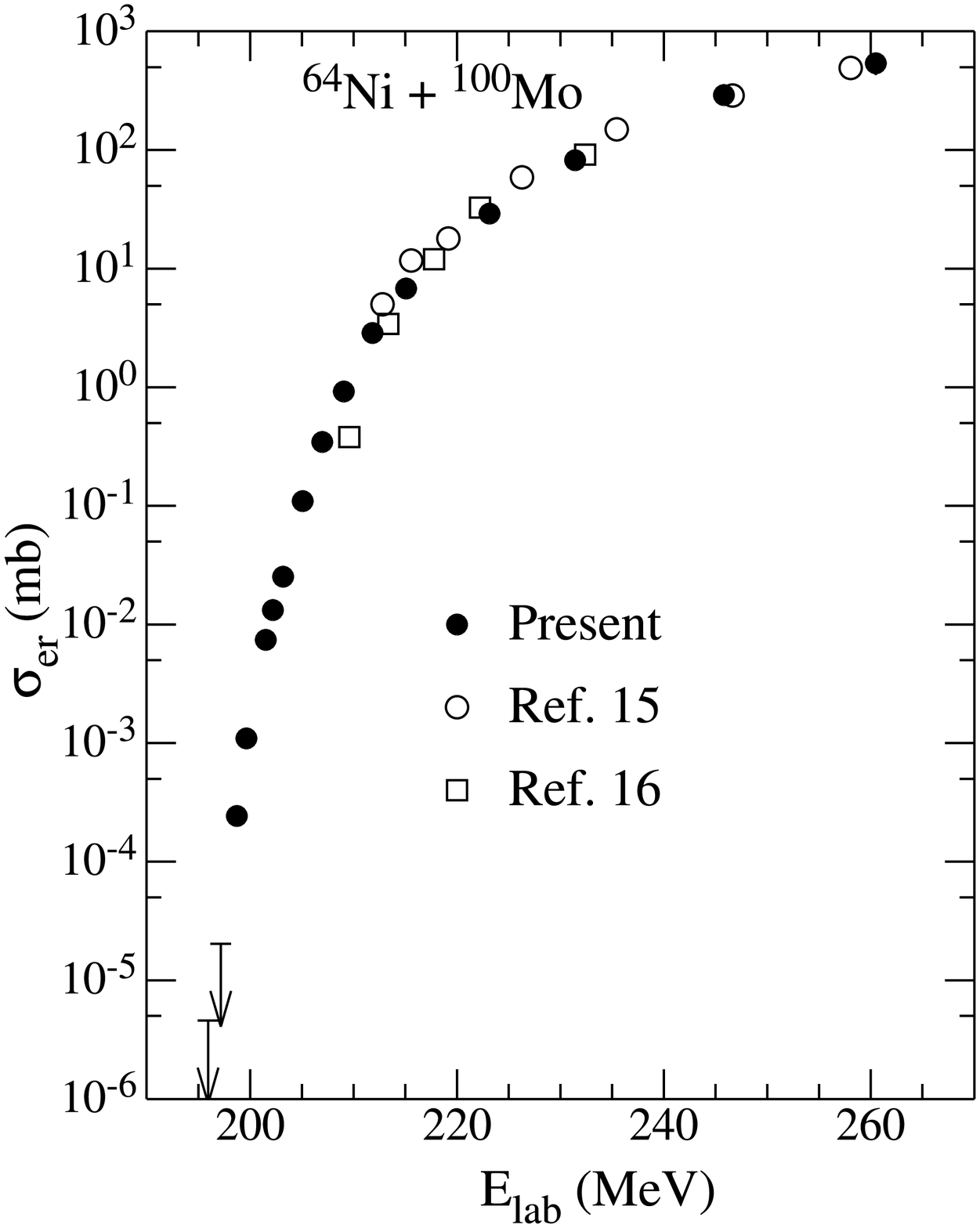,width=8.0cm}
\caption{
Fusion excitation function for the system $^{64}$Ni + $^{100}$Mo (solid circles). Statistical errors are smaller than the symbol size. For the two lowest energies, no residue events have been observed. The upper limits correspond to one count. Included in the figure are two previous measurements from Ref.\cite{rehm} (open circles) and Ref. \cite {halb} (open squares).}
\label{fig5}
\end{figure}

The experimental results for the total fusion cross sections, spanning eight orders of magnitude, are presented as a function of laboratory energy in Fig.~\ref{fig5} (solid circles). The incident energies have been corrected for target thickness and for the steep energy dependence; these corrections are small because rather thin targets were used. 
For the lowest two energies no evaporation residue was observed. The results
are shown as upper limits corresponding to one count in each case. The two 
earlier measurements \cite{rehm,halb} are also shown in Fig.~\ref{fig5} as open
circles and open squares, respectively. Statistical errors are smaller than the size of the symbols. The three measurements are generally in good agreement within the quoted uncertainties, except for the lowest energy point of Ref. \cite{halb}. 

\section{Coupled-channels calculations} 

\begin{table} 
\caption{Measured fusion evaporation cross sections and calculated 
fusion-fission cross sections (using the code CASCADE) for the $^{64}$Ni + 
$^{100}$Mo system. Large uncertainties were assigned to the calculated 
fusion-fission cross sections, leading to rather large errors in the total 
cross sections at the highest beam energies. $N_q$ is the number of charge 
states for evaporation residues which were measured in the experiment.}
\vspace{1mm}
\begin{tabular} {ccccc}
 \hline
 \hline
    $E_{c.m.}$ & $N_q$ & $\sigma_{evap}$ &  $\sigma_{fis}$ & $\sigma_{fus}$\\
       (MeV)   & &   (mb)   &  (mb) & (mb) \\
 \hline
  158.8 & 14 & 264    $\pm$ 35            & 275& 539     $\pm$ 111    \\
  149.9 &  9 & 210    $\pm$ 25            &  80& 290     $\pm$ 42     \\
  141.1 &  4 & 80.0   $\pm$ 8.8           &   2& 82.0    $\pm$ 9.8    \\
  136.1 &  5 & 29.2   $\pm$ 3.0           &   0& 29.0    $\pm$ 3.0    \\
  131.2 &  4 & 6.80   $\pm$ 0.71          &   0& 6.80    $\pm$ 0.71   \\
  129.2 &  4 & 2.87   $\pm$ 0.30          &   0& 2.87    $\pm$ 0.30   \\
  127.5 & 12 & 0.92   $\pm$ 0.10          &   0& 0.92    $\pm$ 0.10   \\
  126.2 &  6 & 0.35   $\pm$ 0.04          &   0& 0.35    $\pm$ 0.04   \\
  125.0 &  4 & 0.109  $\pm$ 0.012         &   0& 0.109   $\pm$ 0.012  \\
  123.9 &  4 & 0.0253 $\pm$ 0.0029        &   0& 0.0253  $\pm$ 0.0029 \\
  123.3 &  8 & 0.0132 $\pm$ 0.0014        &   0& 0.0132  $\pm$ 0.0014 \\
  122.9 &  4 & 7.4$\pm 0.87\times 10^{-3}$& 0& 7.4 $\pm 0.87\times 10^{-3}$ \\
  121.7 & 2 &1.10$\pm 0.16\times 10^{-3}$&  0& 1.10$\pm 0.16\times 10^{-3}$\\
  121.2 & 2 &2.42$\pm 0.41\times 10^{-4}$&  0& 2.42$\pm 0.41\times 10^{-4}$\\
  120.2 & 2 &$<$ 2.0$\times 10^{-5}$ &                 \\
  119.5 & 2 &$<$ 4.6$\times 10^{-6}$ &                 \\
\hline
\end{tabular}
\end{table}

The coupled-channels calculations of Ref. \cite {rehm} reproduced the excitation function quite well down to the 0.5 mb level. It was found that with the same coupling parameters, the calculations cannot reproduce the new data at lower beam energies. In fact, in the literature there are not many such calculations that can reproduce successfully experimental data for heavy systems. Since the nucleus $^{100}$Mo is rather soft, multi-phonon states 
and large coupling effects should be included in the calculations and this represents a challenge.  In the following, two sets of coupled-channels calculations are  presented, including two- and three-phonon states, respectively.

\begin{table} 
\caption{Structure input of low-lying states in $^{64}$Ni and $^{100}$Mo.
For $^{64}$Ni, the B(E$\lambda$)-values for the quadrupole transitions are from 
\cite{char,vide}, 
the octupole strength is from Ref. \cite{brau}.
The $^{100}$Mo input is from \cite{NDSMo}.}
\begin{ruledtabular}
\begin{tabular} {cccccc}
Nucleus &
$\lambda^\pi$ & E$_x$ & B(E$\lambda$) & $\beta_\lambda^C$ & $\beta_\lambda^N$ \\
        &      & (MeV) & (W.u.)&  &\\
\colrule
$^{64}$Ni  & $2^+$   & 1.346 & 8.6 & 0.165 & 0.185\\
        & 2ph($2^+$) & 2.869 & -  & 0.165 & 0.185\\  
           & $3^-$   & 3.560 & 12  & 0.193 & 0.200\\
\colrule
$^{100}$Mo & $2^+$   & 0.536 & 37  & 0.231 & 0.254\\
        & 2ph($2^+$) & 1.002 & 68  & 0.222 & 0.244\\
        & 3ph($2^+$) & 1.671 & 90  & 0.206 & 0.215\\
           & $3^-$   & 1.908 & 32  & 0.210 & 0.230\\
        & 2ph($3^-)$ & 3.816 &  -  & 0.210 & 0.230\\
\end{tabular}
\end{ruledtabular}
\end{table}

\begin{table} 
\caption{Energies and reduced transition probabilities of the I =
0,2 and 4 transitions.}
\begin{ruledtabular}
\begin{tabular} {ccccc}
Nucleus & \multicolumn{2}{c}{$^{64}$Ni} &\multicolumn{2}{c}{$^{100}$Mo}\\
\colrule
$I^\pi$ & $E_I$ & $B(E2,I\to2)$&$E_I$ & $B(E2,I\to2)$\\
          & (MeV) &  (W.u.)      & (MeV) &  (W.u.)\\
\colrule
      $0_2^+$ &    2.867 &110   & 0.695 &  92  \\
      $2_2^+$ &    2.277 &  0   & 1.064 &  51  \\
      $4_1^+$ &    2.610 &$<$37 & 1.136 &  69  \\
Effective-2ph & 2.87-2.75&22-41 & 1.002 &  68  \\
$\frac{<2ph|\alpha _{20} |1ph>}{\sqrt{2/5}}$&-& 0.177-0.242& - & 0.222\\
\end{tabular}
\end{ruledtabular}
\end{table}

The nuclear structure input for the calculations is given in Table II. The structure input for $^{64}$Ni is the same as used in Ref. \cite{jiang2} to analyze the $^{64}$Ni + $^{64}$Ni fusion-evaporation excitation function. The nuclear quadrupole coupling, $\beta_\lambda^N$, for $^{64}$Ni was set $\sim10\%$ higher than the value for Coulomb coupling, $\beta_\lambda^C$, (for reasons discussed in Ref. \cite {vide}). For simplicity, it was assumed that the quadrupole mode corresponds to a perfect vibrator. This is not unreasonable, considering the uncertainty of the predicted one- and two-phonon couplings shown in Table III. The structure input for $^{100}$Mo is also given in Table II. Here, the nuclear $\beta$-values were also set $\sim10\%$ higher than the Coulomb $\beta$-values.

The two-phonon (2ph) calculations include the $2^+$ and $3^-$ one-phonon states in both nuclei, as well as mutual excitations and the two-phonon states listed in Table II. The two-phonon octupole excitation of $^{100}$Mo was also included. The two-phonon states were treated as perfect vibrational states. This gives a total number of 14 coupled channels. The three-phonon (3ph) calculations additionally include all mutual excitations of the states considered in the 2ph calculations up to two mutual two-phonon states. Moreover, the estimated three-phonon excitation of the $2^+$ state in $^{100}$Mo shown in table II was also included. However, states above 5.5 MeV were not included, so the total number of channels is 31. The ion-ion potential was parametrized as a Woods-Saxon well with a depth of $V_0$ = -82.9 MeV, diffuseness $a$ = 0.686 fm, and nuclear radius $R_N$ = 10.19 + $\Delta R$ fm. 

\begin{figure}[hbt] 
\epsfig{file=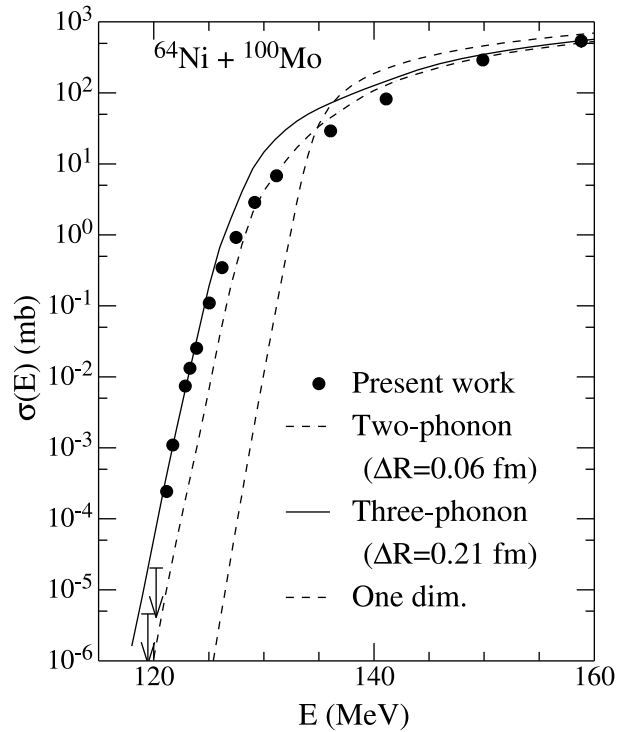,width=8.0cm}
\caption{
Fusion excitation function for $^{64}$Ni + $^{100}$Mo compared with several 
calculations described in the text.}
\label{fig6}
\end{figure}

If the value of $\Delta R$=0.06 fm was used in the two-phonon calculations, the result is similar to the calculations in Ref. \cite {rehm}, and it agrees quite well with the data for the excitation function above the 0.5 mb level. This two-phonon calculation is shown in Fig.~\ref{fig6} as the dotted-dashed curve and it is evident that it does not reproduce the data well at low energies. We have also tried the recipe of a larger diffuseness (up to $a_i$=5 fm) inside the Coulomb barrier, which was described in Ref. \cite {jiang1}, but this did not result in any significant improvement of the fit to the data.

A value of $\Delta R$ = 0.21 fm was needed to reproduce the present data in the cross section region of 0.1 to 100 $\mu$b. The potential with this $\Delta R$ value produces a Coulomb barrier of 134 MeV, and a pocket inside the Coulomb barrier at 112.9 MeV, which is about 20 MeV higher than the ground state energy of the compound nucleus at 92.3 MeV. The value of $\Delta R$ = 0.21 fm was also adopted in the 3ph calculations. The result is shown as the solid curve in Fig.~\ref{fig6}, which is seen to provide a better fit to the data in the 0.1 to 100 $\mu$b cross section range. The dashed curve shows as a reference the results obtained in a one-dimensional calculation, {\it i.e.}, without any couplings (with $\Delta R$ = 0.21 fm). It should be emphasized that the difference between the two sets of calculations, using the same $\Delta R$ = 0.21 fm, is rather modest. Hence, the 2ph calculation with $\Delta R$ = 0.21 fm is not shown for clarity. These two calculations reproduce the experimental data in the highest energy region and in the 0.1 to 10 $\mu$b range, but the cross sections in the region around E=130 MeV are overpredicted. These calculations also overpredict the cross sections at the lowest energies, which is the main topic of the present study.

The coupled-channels calculations shown in Fig.~\ref{fig6} exhibit essentially the same energy dependence (slope) as the one-dimensional calculation when the cross sections are small ($<$ 10 $\mu$b). The coupled-channels calculations are just shifted to lower energies relative to the one-dimensional calculation. In the present case the shift is about 7 MeV (for the same $\Delta R$ = 0.21 fm). This is a general feature of coupled-channels calculations and it is, therefore, very unlikely that any minor adjustment in the
coupled-channels calculations would reproduce the steep falloff that the data exhibit at extreme sub-barrier energies. Thus, it appears that the fusion hindrance behavior, which now has been observed for many systems, is also present in the new data for $^{64}$Ni+$^{100}$Mo. This will be shown more convincingly in the next section, where other 
representations of the fusion cross section are discussed.

From the experimental data and calculations alike, one can infer a rather broad, about 18 MeV wide, barrier distribution (not shown here). An elaborate coupling scheme is therefore needed in the coupled-channels calculations. At energies above the barrier, one-dimensional calculations result in higher cross sections when compared to coupled-channels calculations. This behavior is caused by the long-range Coulomb-excitation/polarization of the low-lying quadrupole states, as pointed out in  Ref. \cite {rehm}. As expected, the suppression of the coupled-channels fusion cross sections is rather strong in this region for the ``soft'' $^{100}$Mo nucleus.

\section{Logarithmic derivatives and $S$-factors} 

\begin{figure}[hbt] 
\epsfig{file=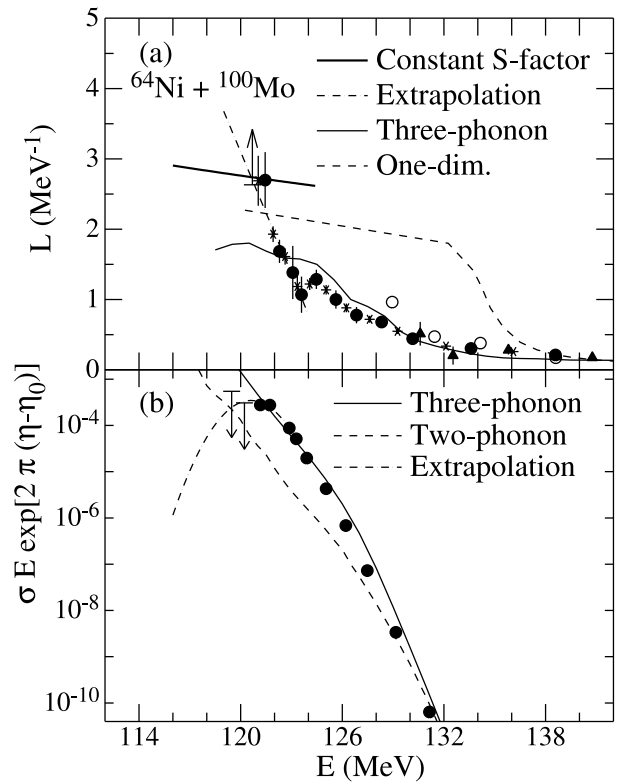,width=8.0cm}
\caption{
(a) Logarithmic derivative representation of the $^{64}$Ni + $^{100}$Mo fusion excitation function. (b) $S$-factor representation of the same data. All the circles were obtained from two successive data points and stars were derived from least square fits to three neighboring data points. Calculations are shown by curves. The extrapolation curve (dashed-dotted) shown in panel (b) was obtained from a  straight line extrapolation of the logarithmic derivative representation in panel (a). Included in (a) are also the data from Ref. \cite{rehm} (solid triangles) and Ref. \cite{halb} (open circles). See text for details.}
\label{fig7}
\end{figure}

The logarithmic derivative, $L=d\ln(\sigma E)/dE$, originally introduced in 
Ref.~\cite{jiang0}, is shown in Fig.~\ref{fig7}a for the $^{64}$Ni + $^{100}$Mo system. The solid circles were obtained directly from two successive data points, whereas the stars were derived from least-squares fits to three neighboring data points. The lower limit of the logarithmic derivative was derived from the upper limit on the cross section at $E_{cm}$=120.2 MeV and the data point at $E_{cm}$=121.2 MeV (see Table I). The present data are compared to those of Refs.~\cite{rehm, halb}, which are represented by open circles and solid triangles in Fig.~\ref{fig7}a, respectively. Only two-point derivatives are shown for these data. The three-phonon coupled-channels and the one-dimensional barrier penetration calculations are shown as solid and dashed curves, respectively, while the thick solid, nearly horizontal line corresponds to a constant $S$-factor expression derived in Ref. \cite {jiang1}. The present experimental results just reach the constant $S$-factor line, implying that the experimental data have reached a maximum value for the $S$-factor. The energy, $E_s$, representing the intersection between the experimental logarithmic derivatives and the constant $S$-factor line corresponds to a value of $E_s$=120.6 MeV. The dashed-dotted line in Fig.~\ref{fig7}a is an extrapolation obtained under the assumption that the logarithmic derivative is a straight line near the crossing point. This method was first introduced in Ref. \cite{jiang2} in order to obtain the extrapolated values of $E_s$ for the systems $^{58}$Ni + $^{60}$Ni and $^{58}$Ni + $^{64}$Ni. The calculated logarithmic derivatives are seen to saturate around $L$=1.5 - 2 MeV$^{-1}$ (or start to oscillate) below $E_{cm}$=124 MeV, whereas the corresponding experimental values continue to grow with decreasing energies. This saturation behavior has already been noted in Refs.~\cite {jiang0,jiang1,jiang2}. 

\begin{table*}[bt] 
\caption{Parameters for Ni-induced fusion reactions on targets
around the Z, N = 28, 40 and 50 shells.
The parameters are: $Z_1Z_2\sqrt{\mu}$, where $\mu=A_1A_2/(A_1+A_2)$ is 
the reduced mass number of the colliding system,
the crossing point energy $E_s$, 
the lowest measured energy $E_{min}$,
the reference energy $E_s^{ref}$ obtained from Eq.~\ref{eq1},
the ratio $E_s/E_s^{ref}$,
the minimum cross section measured,
the fusion reaction $Q$-value, 
the Coulomb barrier (calculated with the Bass model \cite{bass})
and the number of ``valence nucleons'', $N_{ph}$, outside the nearest closed shells.
For systems, which either exhibit a clear maximum in the $S$-factor or whose logarithmic derivatives have not reached the constant $S$-factor curve yet, but can be extrapolated to obtain the crossing point, values $E_s$ are listed in column 3. For systems, where it is not possible to make good extrapolations, the lowest measured energy is quoted in column 4.
For the Ni + Sn systems, where $^{112-124}$Sn targets have been used in the measurements, 
only data for the $^{124}$Sn target are included in the table for brevity.}
\begin{ruledtabular}
\begin{tabular} {cccccccccccc}

System  & $Z_1Z_2\sqrt{\mu}$ & $E_s$ &$E_{min}$& $E_s^{ref}$ & $\frac{E_s}{E_s^{ref}}$&
$\sigma _{min}$&
$- Q$ &$V_c$& $N_{ph}$ & Ref. \\
& & (MeV) & (MeV) & (MeV) & & (mb) &(MeV) & (MeV) & &\\
\tableline
$^{58}$Ni+$^{58}$Ni &4222& 94.0&     &93.0 &1.01 & 0.049 &66.12&102.0& 2+2
&\cite{beck1} \\
$^{58}$Ni+$^{60}$Ni &4258& 92  &     &93.5 &0.98 & 0.040 &62.69&101.3& 2+4
&\cite{stef3} \\
$^{58}$Ni+$^{64}$Ni &4325& 89  &     &94.5 &0.94 & 0.077 &53.04&100.0& 2+4
&\cite{beck2} \\
$^{64}$Ni+$^{64}$Ni &4435& 87.7&     &96.1 &0.91 & $<5.3\times 10^{-6}$&48.78
& 98.1& 4+4 &\cite{jiang2} \\
$^{60}$Ni+$^{89}$Y  &6537&122.9&     &124.5&0.99 & $<9.5\times 10^{-5}$&90.50
&136.5& 4+1 &\cite{jiang0}\\
\tableline
$^{58}$Ni+$^{92}$Mo &7014&     &132.9&130.5&     & 0.17  &108.0&148.6& 2+2
&\cite{rehm} \\
$^{64}$Ni+$^{92}$Mo &7225&     &132.1&133.1&     &  2.6  &100.6&146.0& 4+2
&\cite{rehm} \\
$^{58}$Ni+$^{100}$Mo&7125&     &128.9&131.8&     & 0.72  &90.39&143.3& 2+10
&\cite{rehm} \\
$^{64}$Ni+$^{100}$Mo&7346&120.6&     &134.5&0.90 & $<4.6\times 10^{-6}$&92.29&136.5
& 4+10 & present \\
\tableline
$^{58}$Ni+$^{74}$Ge &5109& 98.5  &     &105.6&0.93 & 0.037 &62.03&113.4& 2+6
&\cite{beck2}\\
$^{64}$Ni+$^{74}$Ge &5249& 97.5  &     &107.5&0.91 & 0.013 &58.48&111.3& 4+6 
&\cite{beck2}\\
\tableline
$^{58}$Ni+$^{90}$Zr &6652&     &127.3&125.9&     & 0.35  &97.24&141.1& 2+0
&\cite{scar}\\
$^{58}$Ni+$^{91}$Zr &6666&     &126.4&126.1&     & 0.25  &94.18&140.8& 2+1
&\cite{scar}\\
$^{58}$Ni+$^{94}$Zr &6708&     &122.7&126.6&     & 0.35  &86.94&139.8& 2+4
 &\cite{scar}\\
$^{64}$Ni+$^{92}$Zr &6881&     &119.0&128.8&     &$7.5\times 10^{-4}$&91.45&137.8
& 4+2 &\cite{jans,stef2,henn}\\
$^{64}$Ni+$^{96}$Zr &6940&     &120.2&129.5&     & 0.15  &86.48&136.6& 4+6
&\cite{stef2}\\
$^{58}$Ni+$^{124}$Sn&8801&     &149.4&151.8&     & 0.19 &112.4&170.3& 2+8 
&\cite{free}\\
$^{64}$Ni+$^{124}$Sn&9096&     &155.0&155.1&     & 23.  &117.5&167.3& 4+8 
&\cite{free}\\
\end{tabular}
\end{ruledtabular}
\end{table*}

The $S$-factor representation for the $^{64}$Ni + $^{100}$Mo system is presented in Fig.~\ref{fig7}b. As the experimental logarithmic derivatives only just reach the constant $S$-factor curve, the $S$-factor maximum is not fully developed. 
Additional measurements would be required to clearly delineate the maximum in the $S$-factor. The dashed-dotted curve corresponds to the straight line extrapolation of the logarithmic derivative in Fig.~\ref{fig7}a. The two- and three-phonon calculations are shown in  Fig.~\ref{fig7}b as dashed and solid curves, respectively.  It is evident that coupled-channels calculations overpredict the fusion cross section at extreme sub-barrier energies.

\section{Comparison to other Ni-induced fusion systems}

\begin{figure}[hbt] 
\epsfig{file=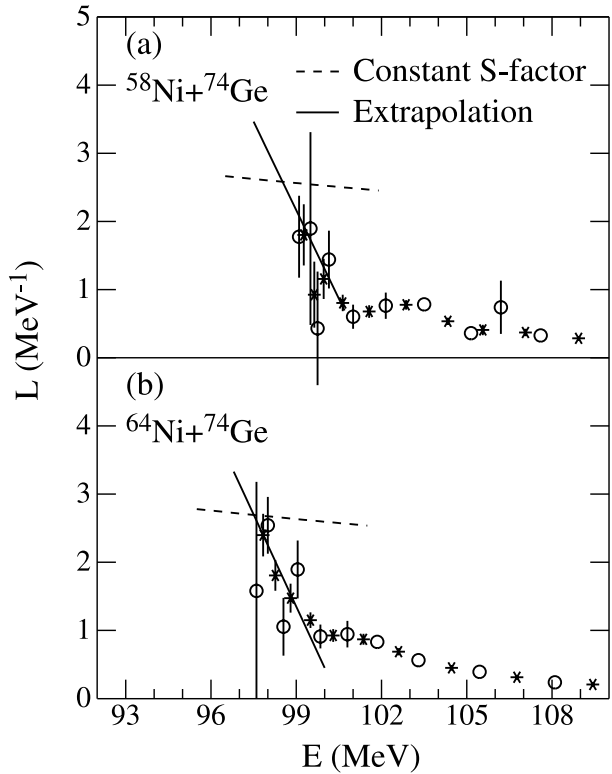,width=8.0cm}
\caption{
Logarithmic derivative representations of the fusion data for the systems
$^{58}$Ni+$^{74}$Ge (a) and
$^{64}$Ni+$^{74}$Ge (b). The data are from Ref. \cite {beck2}.}
\label{fig9}
\end{figure}

The experimental fusion data involving Ni projectiles and compound nuclei in the A=100 - 200 region, are summarized in Table IV. For four systems, namely the $^{58}$Ni+$^{58}$Ni, $^{64}$Ni+$^{64}$Ni, $^{60}$Ni+$^{89}$Y and the present $^{64}$Ni+$^{100}$Mo reactions, the cross sections have been measured to sufficiently low energies to determine the energy, $E_s$ of the maximum of the $S$-factor representation. Previously, we have obtained the $E_s$ values for the $^{58}$Ni+$^{60}$Ni and $^{58}$Ni+$^{64}$Ni systems~\cite{jiang1}. The values of $E_s$ are listed in column three of Table IV. Two additional systems, namely $^{58}$Ni+$^{74}$Ge and $^{64}$Ni+$^{74}$Ge from Ref.\cite{beck2}, have been measured down to levels close to where the $S$-factor maximum occurs. The logarithmic derivative for these systems is shown in Fig.~\ref{fig9}. The location of the $S$-factor maximum was obtained by performing a small extrapolation (solid line) of the logarithmic derivative to where it crosses the constant $S$-factor line (dashed line). The resulting values are listed in Table IV. Because of the uncertainties inherent in these extrapolations, errors of 2$\%$ ($\sim$2 MeV) were assigned to the extrapolated $E_s$ data. Table IV also lists other fusion systems involving Ni beams for which the cross section was not measured down to a level that allows for determination of the $S$-factor maximum. For these systems only the lowest measured energy is listed. 

\begin{figure}[bt] 
\epsfig{file=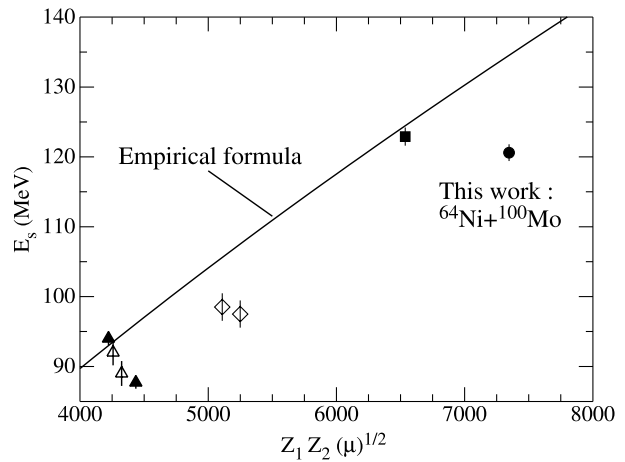,angle=270,width=8.0cm}
\caption{
Plot of $E_s$ vs. $Z_1Z_2\sqrt{\mu}$ for Ni bombarding different targets
(see Table IV). Solid symbols correspond to systems for which the $S$-factor maximum is well determined:
$^{58}$Ni+$^{58}$Ni,
$^{64}$Ni+$^{64}$Ni,
$^{60}$Ni+$^{89}$Y
and $^{64}$Ni+$^{100}$Mo.
Open symbols are associated with the extrapolations for the systems
$^{58}$Ni+$^{60}$Ni,
$^{58}$Ni+$^{64}$Ni,
$^{58}$Ni+$^{74}$Ge and
$^{64}$Ni+$^{74}$Ge.
The triangles represent Ni+Ni, diamonds Ni+Ge, a square
$^{60}$Ni+$^{89}$Y and a circle $^{64}$Ni+$^{100}$Mo, respectively.}
\label{fig8}
\end{figure}

The values of $E_s$ are plotted as a function of the parameter $Z_1Z_2\sqrt{A_1A_2/(A_1+A_2)}$ in Fig.~\ref{fig8} and compared to the empirical formula (solid curve), Eq.~\ref{eq1}, obtained from a fit to all available fusion data involving stiff nuclei. Obviously, only two systems, $^{58}$Ni+$^{58}$Ni and $^{60}$Ni+$^{89}$Y follow the systematics; all other systems fall below the curve. Previously, it has been pointed out~\cite{jiang2} that  there is a rather compelling correlation between the stiffness of the interacting nuclei and the location of the $S$-factor maximum, $E_s$, relative to the empirical trend for the Ni+Ni systems. The addition to the systematics of the data points for $^{64}$Ni+$^{100}$Mo and for the two Ni+Ge systems, which all involve soft nuclei, appears to corroborate this observation. Thus, for the $^{64}$Ni+$^{100}$Mo system, the value of $E_s^{ref}$ predicted from Eq.~\ref{eq1} is $E_s^{ref}$=134.5 MeV whereas the measured value of $E_s$=120.6 MeV is only 90$\%$ of this value. The reason for this reduction is presumably that $^{64}$Ni + $^{100}$Mo should be viewed as an open-shell colliding system so that strong coupling effects broaden the effective barrier distribution and push the energy where the hindrance behavior occurs down to even lower energies.

\begin{figure}[bt] 
\epsfig{file=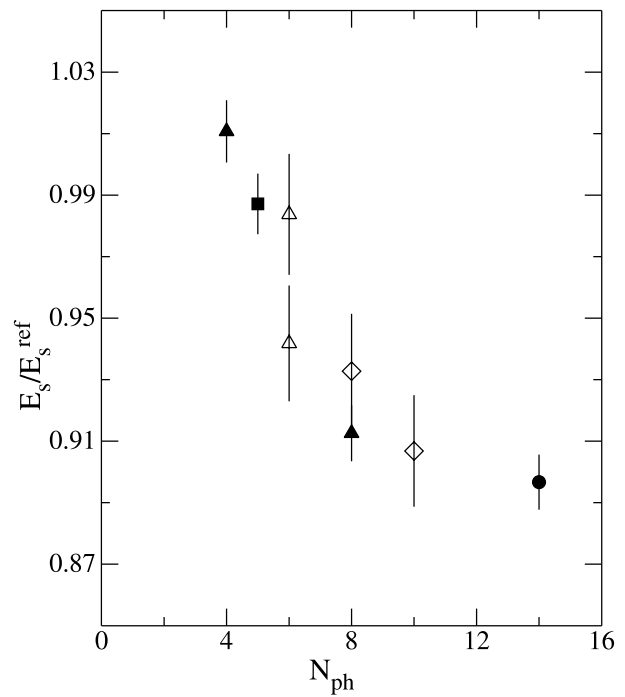,width=8.0cm}
\caption{Plot of $E_s/E_s^{ref}$ vs. $N_{ph}$, where $N_{ph}$ is the total number of
  ``valence nucleons'' outside closed shells in the entrance channel.
Symbols are defined in Fig.~\ref{fig8}.}
\label{fig10}
\end{figure}

The deviation of the measured or extrapolated values of $E_s$ from the expected $E_s^{ref}$ value, seen in Fig.~\ref{fig8}, thus appears to depend on the stiffness of the fusing nuclei. A quantitative relation between the stiffness and the deviation from $E_s^{ref}$ is not yet known. As a first attempt, we associate the stiffness of a nucleus to its proximity to closed proton or neutron shells and define the number of ``valence nucleons'', $N_{ph}$, as the sum of particles and holes outside the nearest closed shells. Here, $^{64}$Ni is considered to have four holes in the N=40 neutron shell rather than eight particles outside N=28. The values of $N_{ph}$ are listed in Table IV and, in Fig.~\ref{fig10}, the ratio $E_s/E^{ref}_s$ is plotted as a function of this parameter. There is a general trend of decreasing values of $E_s/E_s^{ref}$ with increasing values of $N_{ph}$. We observe that for the other systems in Table IV, the data for $E_{min}$ (upper limits) are not in contradiction with Figs.~\ref{fig8} and \ref{fig10}. It should be noted that all of the systems shown in Figs.~\ref{fig8} and \ref{fig10} and Table IV have rather large negative fusion $Q-$values. One may also compare the $E_s$-values to the height of the Coulomb barrier (obtained from the Bass prescription), which is listed in Table IV. The ratio $E_s/E_s^{ref}$ exhibit a stronger dependency on the value of $N_{ph}$ than that obtained for the ratio $E_s/V_c$ indicating that the observed effect does not just depend on the change in the Coulomb barrier height with addition of neutrons to the interacting nuclei.

\section{Discussion}

The above observations are all phenomenological in nature as there is at present no satisfactory understanding of the fusion hindrance at extreme sub-barrier energies. Many authors have tried to explain this new phenomenon \cite{lin,linr,hagi,dass,gira,proc1,proc2}. One suggestion is to use a large diffuseness of the ion-ion potential in the coupled-channels calculations. This recipe is sometimes used to explain high-precision fusion data \cite{hagi}. It has been argued that the failure to reproduce the steep fall-off is caused by the Hill-Wheeler approximation \cite{lin, linr}. 
In Ref. \cite{jiang1} it was shown, however, that these limitations of the analysis are not responsible for the observed sub-barrier fusion hindrance phenomenon. Dasso and Pollarolo tried to reduce the fusion cross sections by using a shallow well inside the barrier \cite{dass}. Giraud et al. investigated the effect of ``friction'' \cite {gira}. More discussions about this phenomenon can be found in Refs.~\cite{proc1,proc2}. All of these suggestions may improve the fit of calculations to the experimental data in some cases, but they do not provide a convincing explanation of the observed suppression in all systems. These studies are presently at the stage of raising questions and discovering weak points in the existing models. The data are most likely still insufficient to lead to the correct explanation. More precision sub-barrier fusion measurements are required to further explore  which modifications in the theoretical models are relevant for a correct description of the phenomenon. 

\section{Summary and conclusion}

The phenomenon of sub-barrier fusion hindrance was first observed in systems involving stiff nuclei and a simple expression~\cite{jiang0,jiang1} was derived for the energy at which the hindrance of fusion between such nuclei occurs. Furthermore, a study of Ni+Ni fusion involving different Ni isotopes~\cite{jiang2} has shown that the onset of the fusion hindrance deviates from these systematics depending on the ``stiffness'' of the interacting nuclei.

In the present work, we have measured the fusion excitation function for the system $^{64}$Ni+$^{100}$Mo down to a cross section level of $\sim$5 nb, {\it i.e.} about 12\% below the Bass barrier. We observe that the fusion process is hindered at the lowest energies relative to expectations based on coupled-channels calculations. The present study of the $^{64}$Ni+$^{100}$Mo system shows that the fusion hindrance for this system, as well as two other soft systems, deviates strongly from the systematics, and it thus corroborates the earlier observation in Ni+Ni systems. It is furthermore shown that this deviation depends monotonically on a parameter $N_{ph}$, which is the sum of nucleons (holes) outside of closed shells of the fusing nuclei suggesting that this parameter is a good measure of the ``stiffness'' of the interacting nuclei.

In conclusion, an interesting nuclear structure dependence of the fusion hindrance has been observed. The origin of this effect is still unknown. It occurs at relatively high excitation energies (for systems in Fig.~\ref{fig8} and \ref{fig10}, they are around $E_{ex} \sim$ 30 - 40 MeV), where the natural width of compound levels is larger than their spacing. A lack of available final states thus appears to be ruled out as an explanation.

{\bf Acknowledgments}
This work was supported by the US Department of Energy, Office of Nuclear
Physics, under Contract Nos. W-31-109-ENG-38 and DE-FG-06-90ER-41132.


\begin{references}
\bibitem{sum1} M. Beckerman, Physics Report, {\bf 129}, 145 (1985). 
\bibitem{sum2} M. Beckerman, Rep. Prog. Phys. {\bf 51}, 1047 (1988).
\bibitem{sum3} R. Vandenbosch, Annu. Rev. Nuc. Part. Sci. {\bf 42}, 447 (1992).
\bibitem{sum4} A. B. Balantekin and N. Takigawa, Rev. Mod. Phys. {\bf 70},
  77 (1998).
\bibitem{sum5} M. Dasgupta, D.J. Hinde, N. Rowley and A.M. Stefanini, 
  Annu. Rev. Nucl. Part. Sci. {\bf 48}, 401 (1998).
\bibitem{hagi0} K. Hagino and N. Takigawa, Phys. Rev. C {\bf 55}, 276 (1997).
\bibitem{rowl} N. Rowley, G.R. Satchler and P.H. Stelson, Phys. Lett.
  {\bf B254}, 25 (1991).
\bibitem{jiang0} C.L. Jiang {\it et al.,} Phys. Rev. Lett. {\bf 89}, 052701 
  (2002).
\bibitem{jiang1} C.L. Jiang, H.Esbensen, B.B. Back, R.V.F. Janssens and
  K.E. Rehm, Phys. Rev. C. {\bf 69}, 014604 (2004).
\bibitem{jiang2} C.L. Jiang {\it et al.,} Phys. Rev. Lett. {\bf 93}, 012701 
  (2004).
\bibitem{beck1} M. Beckerman, J. Ball, H. Enge, M. Salomaa, A. Sperduto,
  S. Gazes, A. DiRienzo and J.D. Molitois, Phys. Rev. C {\bf 23}, 1581 (1982).
\bibitem{stef3} A.M. Stefanini {\it et al.,} Phys. Rev. Lett. {\bf 74}, 864
  (1995).
\bibitem{beck2} M. Beckerman {\it et al.,} Phys. Rev. C {\bf 25}, 837 (1982).
\bibitem{acke} D. Ackermann {\it et al.,} Nucl. Phys. {\bf A609}, 91 (1996).
\bibitem{rehm} K.E. Rehm, H. Esbensen, J. Gehring, B. Glagola, D. Henderson,
  W. Kutschera, M. Paul, F. Soramel and A.H. Wuosmaa,
  Phys. Lett. {\bf B317}, 31 (1993).
\bibitem{halb} M.L. Halbert, J.R. Beene, D.C. Hensley, K. Honkanen, T.M.
  Semkow, V. Abenante, D.G. Sarantites, and Z. Li,
  Phys. Rev. C. {\bf 40}, 2558 
  (1989).
\bibitem{dav1} C.N. Davids and J.D. Larson, Nucl. Instr. and Meth. 
  {\bf B40/41}, 1224 (1989);
  C.N. Davids, B.B. Back, K. Bindra, D.J. Henderson, W. Kutschera, T.
  Lauritsen, Y. Nagame, P. Sugathan, A.V. Ramayya and W.B. Walters,
  Nucl. Instr. and Meth. {\bf B70}, 358 (1992).
\bibitem{dav3} C.N. Davids, 
  ANL Phys. Div. Annual Report {\bf ANL-03-23}, (2003), p. 104.
\bibitem{hend} D.G. Kovar and D.J. Henderson, 
  ANL Phys. Div. Annual Report {\bf ANL-90-18}, (1990), p. 108.
\bibitem{penn} T.O. Pennington, D.J. Henderson, D.Seweryniak, K.E. Rehm,
  C.L. Jiang, C.N. Davids, C.J. Lister, B.J. Zabransky and B. Blank,
  ANL Phys. Div. Annual Report {\bf ANL-03-23}, (2003), p. 105.
\bibitem{jiang3} C.L. Jiang {\it et al.,} to be submitted to Nuclear Instrument
  and Methods A.
\bibitem{pace}  A. Gavron, Phys. Rev. C {\bf 21}, 230 (1980).
\bibitem{jian} C.L. Jiang and C.N. Davids, ANL Phys. Div. Annual Report {\bf
  ANL-95/14}, (1995), p. 74.
\bibitem{puhl} F. P\"{u}hlhofer, Nucl. Phys. {\bf A280}, 267 (1977).
\bibitem{char} J. Charbonneau, N.V. De Castro Faria, J. L'Ecuyer and 
  D. Vitoux, Bull. Am. Phys. Soc. {\bf 16}, 625 (1971).
\bibitem{vide} F. Videbaek, P.R. Christensen, O. Hansen and K. Ulbaek,
  Nucl. Phys. {\bf A256}, 301 (1976).
\bibitem{brau} M. R. Braunstein, J.J. Kraushaar, R.P. Michel, J.H.
  Mitchell, R.J. Peterson, H.P. Blok, H. de Vries, Phys. Rev. C {\bf 37}, 
  1870 (1988).
\bibitem{NDSMo} B. Singh, At. Data Nucl. Data Tables {\bf 81}, 1 (1997).
\bibitem{hagi} K. Hagino, N. Rowley and M. Dasgupta,
  Phys. Rev. C {\bf 67}, 054603 (2003).
\bibitem{bass} R. Bass, Nucl. Phys. {\bf A231}, 45 (1974). 
\bibitem{scar} F. Scarlassara, S. Beghini, F. Soramel, S. Signorini, L. 
  Corradi, G. Montagnoli, D.R. Napoli, A.M. Stefanini and Z.C. Li,
  Z. Phys. A {\bf 338}, 171 (1991).
\bibitem{stef2} A.M. Stefanini, L. Corradi, H. Moreno, L. M\"{u}ller, D.R.
  Napoli, P. Spolaore and E. Adamides, Phys. Lett. B {\bf 252}, 43 (1990).
\bibitem{jans} R.V.F. Janssens, R. Holzmann, W. Henning, T.L. Khoo, K.T.
  Lesko, G.S.F. Stephans, D.C. Radford and A.M. Van Den Berg, 
  Phys. Lett. B {\bf 181}, 16 (1991).
\bibitem{henn} W. Henning {\it et al.}, private communication (1994).
\bibitem{free} W.S. Freeman, H. Ernst, D.F. Geesaman, W. Henning, T.J.
  Humanic, W. K\"{u}hn, G. Rosner, J.P. Schiffer, B. Zeidman and F.W. Prosser,
  Phys. Rev. Lett {\bf 50}, 1563 (1991).
\bibitem{lin} C. J. Lin, Phys. Rev. Lett. {\bf 91}, 229201 (2003).
\bibitem{linr} C. L. Jiang {\it et al.}, Phys. Rev. Lett. {\bf 91},
  229202 (2003).
\bibitem{dass} C.H. Dasso and G. Pollarolo, Phys. Rev. C {\bf 68}, 054604
  (2003).
\bibitem{gira} B.G. Giraud, S. Karataglidis, K.Amos and B.A. Robson,
  Phys. Rev. C {\bf 69}, 064613 (2004).
\bibitem{proc1} Procedings of the International Conference Fusion03: From a
  Tunneling Nuclear Microscope to Nuclear Processes in Matter, 
  Nov. 12-15, 2003, Matsushima, Japan, Prog. Theor. Phys. Suppl. No. 154, 2004.
  D.J. Hinde, p. 1;
  H. Esbensen, p. 11;
  C.L. Jiang, p.61;
  C.J. Lin, p. 184;
  G. Pollaroro, p. 201;
  M. Dasgupta, p. 209;
  D. Brink, p. 268; 
  A.B. Balantakin, p. 465.
\bibitem{proc2} S.V.S. Sastry {\it et al.}, Conference Book of International
  Conference Fusion03: From a Tunneling Nuclear Microscope to Nuclear 
  Processes in Matter, Nov. 12-15, 2003, Matsushima, Japan, p. 143, 2003.
\end{references}
\end{document}